\documentclass[11pt,twoside]{article}
\usepackage{asp2004}
\usepackage{psfig}
\usepackage{epsf}
\usepackage{graphicx}
\usepackage{graphics}
\usepackage{lscape}
\def\approxgt{\mathrel{\hbox{\rlap{\lower.55ex \hbox {$\sim$}}
        \kern-.3em \raise.4ex \hbox{$>$}}}}
\def\approxlt{\mathrel{\hbox{\rlap{\lower.55ex \hbox {$\sim$}}
        \kern-.3em \raise.4ex \hbox{$<$}}}}
\def\edcomment#1{\iffalse\marginpar{\raggedright\sl#1\/}\else\relax\fi}
\reversemarginpar
\begin{document}
\title{The X-ray variability properties of PG quasars: XMM-Newton results}
\author{M. Guainazzi, E.Jim\'enez-Bail\'on, E.Piconcelli}
\affil{European Space Astronomy Center of ESA, Apartado 50727, E-28080, Madrid, Spain}

\begin{abstract}
This paper presents preliminary results of a systematic study on the X-ray
spectral variability of PG quasars with XMM-Newton. We
concentrate on those objects, whose X-ray
spectra are well fit in the framework of the ``double
Comptonization'' model. On short
($\Delta t \sim 10^3$~s)
timescales, variability between the soft ($E < 2$~keV) and the
hard ($E > 2$~keV) bands is uncorrelated. In the two
objects, for which observations at different epochs
are available, spectral
variability in the soft X-ray regime is unveiled.
Its timescale is constrained
in the range between $\sim$1 week and $\sim$1 year. 
\end{abstract}
\thispagestyle{plain}

\section{Introduction}

In this paper we present preliminary results of a
systematic study on the X-ray spectral
variability properties of the PG quasar sample \cite{schmidt83}
with XMM-Newton \cite{jansen01}. The main goals
of this projects are:

\begin{itemize}

\item the identification of the physical driver underlying spectral variability
in Active Galactic Nuclei (AGN), by determining the typical timescales
on which
it occurs in X-ray unobscured AGN

\item the study of the correlation between emission
in the soft ($E < 2$~keV) and
hard ($E \ge 2$~keV) energy bands.
The results of this study can shed some light on the physical
origin of the X-ray emission. While the hard X-ray emission is probably
due to Comptonization of hard disk soft photons in a compact corona within
a few Schwarzschild radii from the supermassive black hole \cite{haardt91},
the nature of a prominent and almost ubiquitous ``soft excess'' is still
matter of debate. Thermal emission from the disk \cite{czerny87}, 
bremsstrahlung emission from the ionized skin of the accretion disk
(Nayakshin, Kazanas \& Kallman 2000),
Comptonization by cool, thick gas \cite{obrien01},
or an extremely relativistic warm outflow \cite{gierlinski04} are
among the possible
explanations.

\end{itemize}

It is of paramount importance that X-ray variability studies are
performed with the same instrument, as the cross-calibration
among detectors flown on different mission is not yet accurate enough
to guarantee against spurious results. XMM-Newton is ideally
suitable for such a study,
thanks to its unprecedented effective area in the whole 0.1--15~keV energy
band, which allows to expand 
the pioneering studies conducted by RXTE
(Markovitz et al. 2003; McHardy et al. 2004) to a larger
sample of comparatively weaker AGN.

\section{The sample}

Our parent sample is the whole set of XMM-Newton observations of PG quasars
available in the public archive as of August 2004. They cover 42 objects, with
largely inhomogeneous exposure times. Their spectral
properties are presented by Porquet et al. (2004)
and Jim\'enez-Bail\'on et al. (2004). In this paper we
will concentrate on the spectral variability pattern in objects
dominated by Comptonization in the whole
XMM-Newton energy bandpass (0.1--10~keV).
Consequently, we will consider in the following only 
the 20 PG quasars,
whose XMM-Newton EPIC spectra are well fit with a double power-law
model by Jim\'enez-Bail\'on et al. (2004).

\subsection{Short term ($<$1~day) variability}

\begin{figure}
\centering
\hbox{
\hspace{-1.0cm}
\includegraphics[width=7cm]{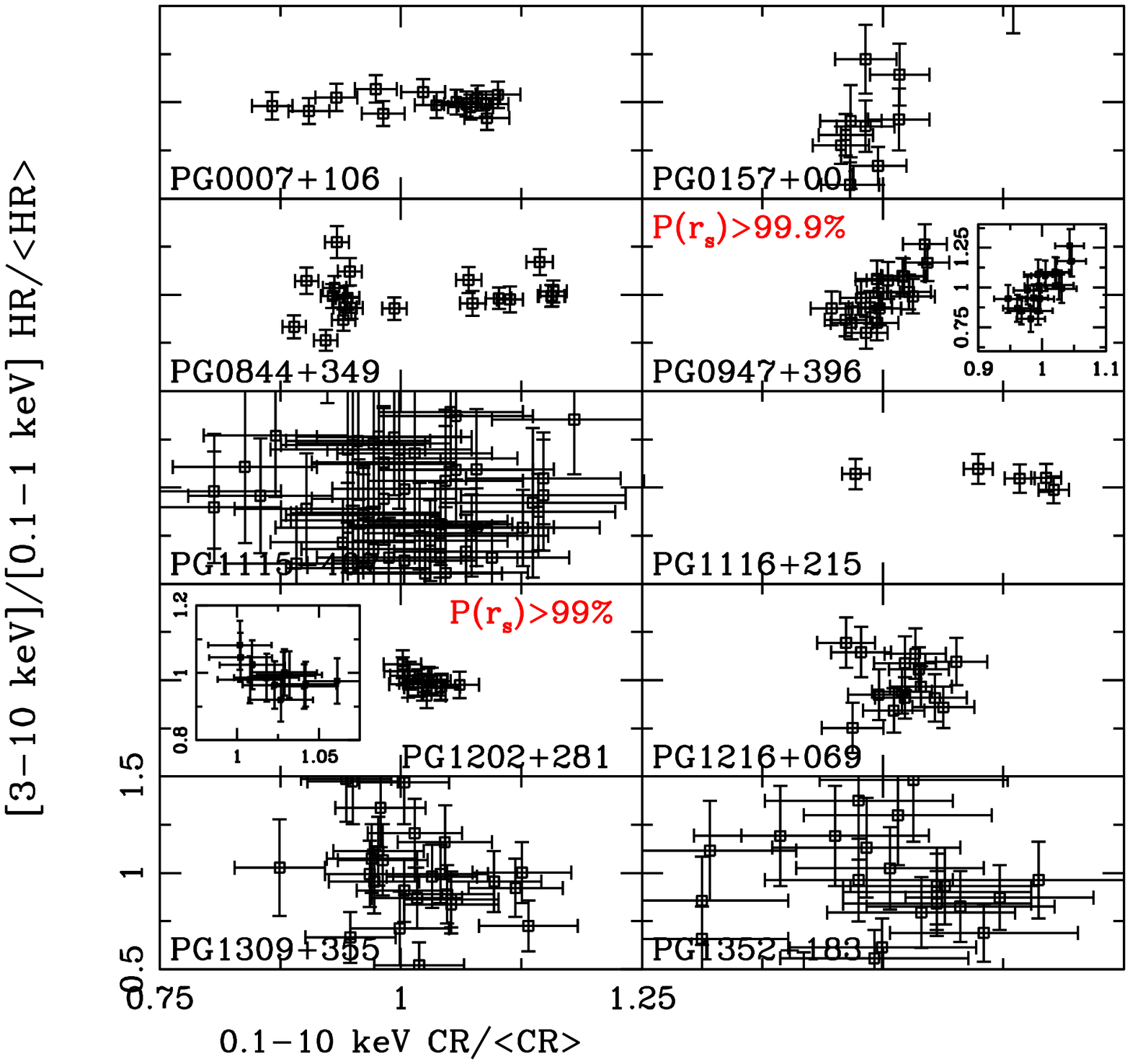}
\includegraphics[width=7cm]{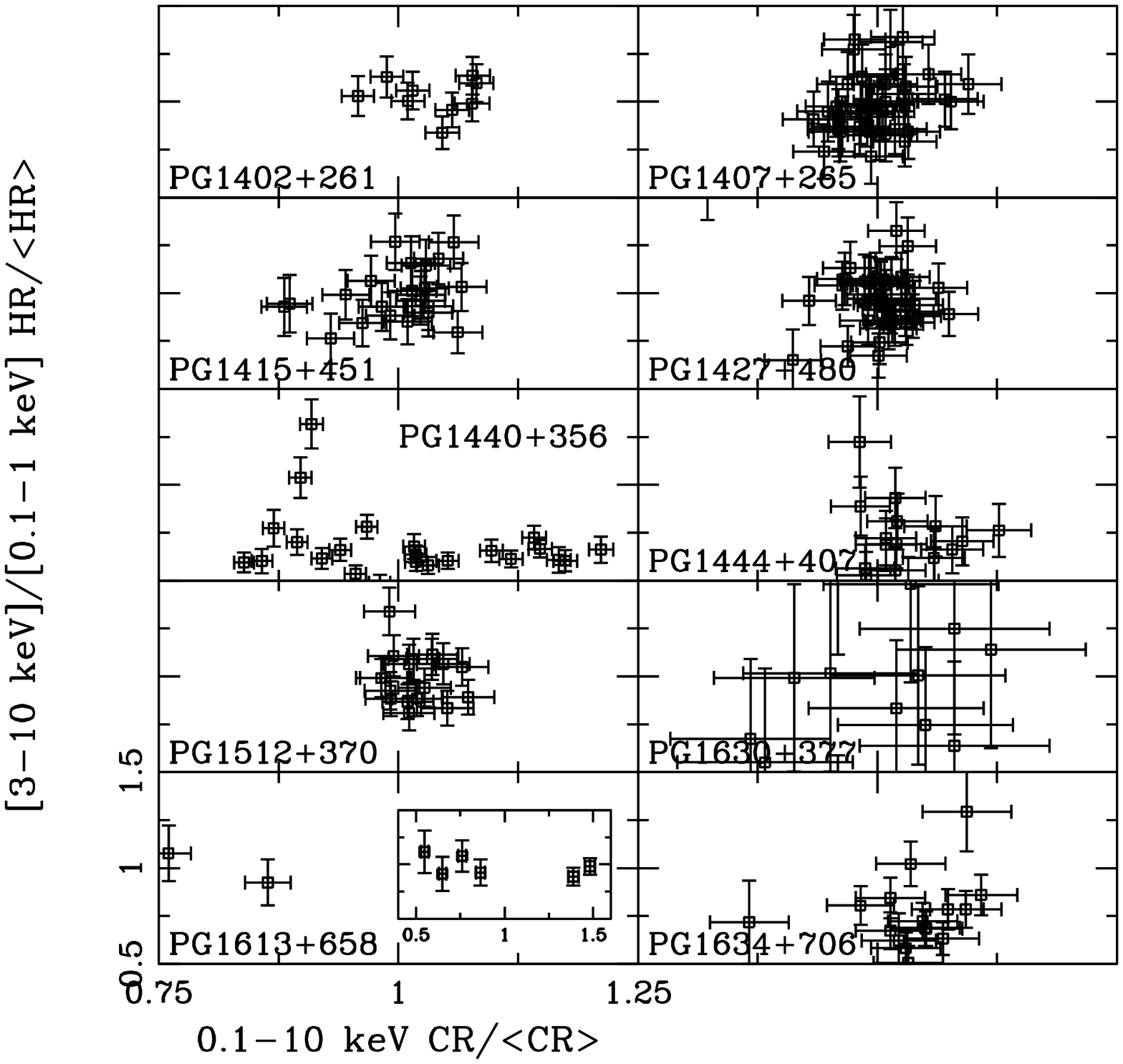}
}
\caption{3--10~keV/0.1--1~keV Hardness Ratio (HR) versus 
0.1--10~keV count rate for the PG quasars observed by XMM-Newton,
whose spectrum can be fit in the ``double Comptonization'' scenario.
The cases where a (positive or negative) correlation is found at
a confidence level larger than 99\% are highlighted
              }
\label{fig1}
\end{figure}

In Fig.~\ref{fig1} we show the 3--10~keV versus 0.1--1~keV hardness ratio (HR)
against the 0.1--10~keV count rate as measured on light curves with
a time binning $\Delta t = 1024$~s. In only
2 objects ($\simeq$10\% of the cases) one
observes a correlation more significant
then the 99\% confidence level according to a K-S test. Moreover,
these correlations have two opposite signs: positive in
PG0947+396, and negative in PG1202+281. We conclude that on
such short timescales
variations of the X-ray emission produced by the two Comptonization media are
largely independent. They are probably dominated by local
fluctuations in the accretion flow or in the disk illumination, which do not
affect the disk-corona system as a whole.

\subsection{Long-term ($>$1 day) variability)}

\begin{figure}
\centering
\hbox{
\hspace{-1.0cm}
\psfig{figure=guainazzi_fig3.ps,height=6.0cm,width=7.0cm,angle=-90}
\psfig{figure=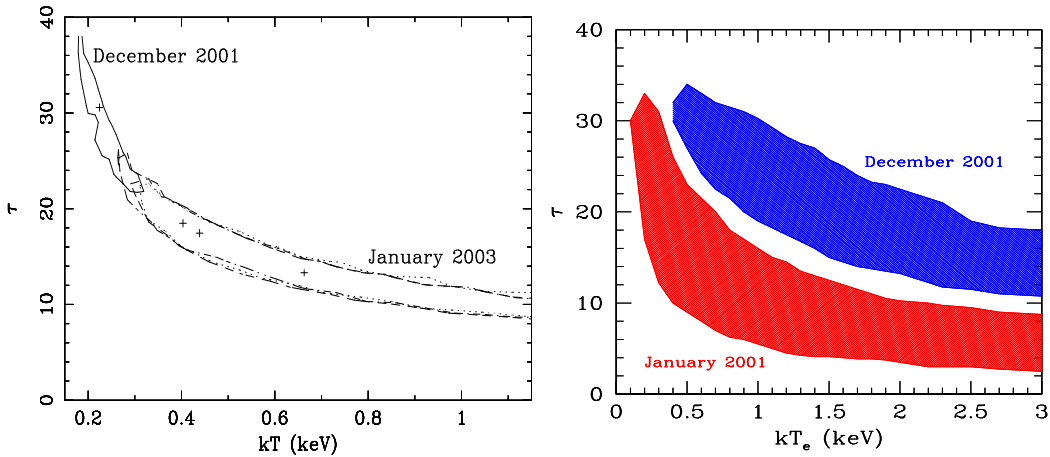,height=6.0cm,width=7.0cm}
}
\caption{Iso-$\chi^2$ contour plots for the temperature
versus the optical depth in the Comptonized plasma responsible
for the soft X-ray emission in the XMM-Newton observations
of PG1440+256 ({\it left}) and PG1407+265 ({\it right}).
The outermost contours correspond in both
cases at the 99\% confidence level for 2 interesting
parameters.
              }
\label{fig2}
\end{figure}

There are only two PG quasars, for which XMM-Newton observations
at multiple epochs
are available: PG1407+265 and PG1440+356.
The former was observed twice, in January and December 2001.
The latter was observed by XMM-Newton 4 times
in the framework of a specific program
to determine the
timescale of the spectral variability pattern. The
first observation was performed on December 2001, the remaining three in
January 2003, at the relative distance of three days each. In order to
spectrally characterize their variability properties, we fit the data
with a combination of two Comptonizing continua, trying different
geometries for the seed photons influx and the Comptonizing
media. The following results refer to a physical scenario
where the two spherical reservoirs of Comptonizing electrons
(a two-phase corona) share
the same soft photons input flux (a region 
at the peak of the accretion disk thermal distribution,
and smaller than the disk gradient temperature scale).
We show in Fig.~\ref{fig2}
the iso-$\chi^2$ contour plot for the temperature ($kT$) versus
plasma optical depth ($\tau$) for the Comptonizing medium producing
the bulk of the X-ray emission in the soft X-ray band. In both cases the
best-fit parameters
corresponding to observations separated by the longer intervals
($\simeq$1~year) are different at a confidence level
larger then 99\% for two interesting parameters.
Intriguingly enough, in PG1407+265
a change in the optical depth of the Comptonized medium
is favored, whereas in PG1440+356 a feedback mechanism
between these quantities may force the contours to follow a
source-specific ``track'' in the kT vs. $\tau$ plane.
Given the well known
degeneracy of the above parameters in Comptonization models
\cite{brinkmann04}, one
should refrain from attributing a too literal meaning to the best-fit
parameters values. However, regardless of their true values,
one can firmly conclude that in these two
cases the mechanism responsible for the spectral variability
operates on time scales longer than about 1~week and shorter than
about 1~year. Although this is still a rather loose constraint
(which could be improved in the future if XMM-Newton will pursue
similar monitoring programs sampling intermediate timescales), it
already rules out dynamical timescales playing an important role.
Viscous instabilities in the accretion flow are a more likely
candidate for the ultimate physical driver of the long-term
variability.

\section{Conclusions}

We are undertaking a systematic study of the X-ray variability pattern
of the PG quasars using XMM-Newton observations. In this paper we focus
our attention on the variability pattern in ``naked'' quasars, whose
emission across the whole XMM-Newton band is due to Comptonization.
The main results can be summarized as follows:

\begin{itemize}

\item on short ($\approxlt$hours) timescales there is no correlation
between the variability in the soft and the hard X-ray regime

\item variability of the Comptonizing plasma physical parameters is
observed when observations separated by more than 1 week (and less
than 1 year) are compared

\end{itemize}

We propose an interpretation in terms of viscous instabilities in the
accretion flow driving the observed changes
in the properties of the Comptonizing
plasma.

\end{document}